# Subcellular Metabolic Tracking Using Fluorescent Nanodiamonds Relaxometry

*Jia Su†, Linyu Zeng†, Pengyu Chen, Zenghao Kong, Fazhan Shi\**

J. Su, L. Zeng, P. Chen, Z. Kong, F. Shi

Laboratory of Spin Magnetic Resonance, School of Physical Sciences, Anhui Province Key Laboratory of Scientific Instrument Development and Application, University of Science and Technology of China, Hefei 230026, China

E-mail: fzshi@ustc.edu.cn.

L. Zeng, F. Shi

Hefei National Research Center for Physical Sciences at the Microscale, University of Science and Technology of China, Hefei 230026, China

Z. Kong, F. Shi

Hefei National Laboratory, University of Science and Technology of China, Hefei 230088, China

F. Shi

School of Biomedical Engineering and Suzhou Institute for Advanced Research, University of Science and Technology of China, Suzhou 215123, China

F. Shi

The First Affiliated Hospital of USTC, Division of Life Sciences and Medicine, University of Science and Technology of China, Hefei 230026, China

†These authors contributed equally to this work

Funding: This work was supported by the National Natural Science Foundation of China (Grant No. T2125011), the Chinese Academy of Sciences (Grant No. YSBR-068), Innovation Program for Quantum Science and Technology (Grant No. 2021ZD0302200, 2021ZD0303204), New Cornerstone Science Foundation through the XPLORER PRIZE, and the Fundamental Research Funds for the Central Universities.

**Abstract**

Fluorescent nanodiamonds (FNDs) relaxometry holds promising future for advancement of high spatiotemporal resolution metabolic imaging technology. In this study, we demonstrate a simultaneous integration of spatial position tracking with FND relaxometry to characterize the temporal dynamics of metabolic processes, thereby enhancing the capability to monitor cellular activities over time. To enable targeted metabolic probing in living cells, FNDs were surface-functionalized to achieve specific localization within key organelles, including the nucleus and mitochondria. This strategy not only facilitates subcellular-level metabolic monitoring but also allows for direct comparison between intra- and extranuclear microenvironments within the same living cell, offering substantial potential for elucidating the spatial and functional heterogeneity of cellular metabolism.

## 1. Introduction

Fluorescent nanodiamond (FND) quantum sensing represents an attractive advancement in nanoscale metrology and biological imaging.[1–5] FNDs—particularly those incorporating Nitrogen-Vacancy (NV) centers,[6,7] atomic-scale defects formed by a nitrogen atom adjacent to a vacancy in the diamond lattice—have emerged as highly sensitive probes capable of detecting ultraweak magnetic fields,[8,9] minute temperature gradients,[10] and localized biochemical fluctuations.[11] NV centers possess intrinsic quantum mechanical properties (electron spin S = 1), enabling optical polarization and readout of spin states.[12] This capability confers exceptional capability for detecting weak magnetic fields and allows for real-time, nanoscale-resolution monitoring of dynamic processes within living cells, offering a powerful advantage over classical sensing modalities.[2,13,14]

The principal advantage of FND-based relaxometry is exquisitely sensitive to the local microenvironment, particularly to paramagnetic species including free radicals—most notably reactive oxygen species (ROS)[15]—as well as paramagnetic metal ions such as iron, gadolinium, and manganese.[16–20] The resulting changes in spin-lattice relaxation time ($T_1$) provide a quantitative measure of these environmental factors. Importantly, because $T_1$ relaxometry is selectively responsive to paramagnetic species such as ROS, it functions as a direct, quantitative reporter of redox activity. Variations in $T_1$, frequently observed in intracellular environments, reflect elevated magnetic noise and serve as sensitive indicators of organelle activity and alterations in the intracellular microenvironment. Conventional organic fluorescent molecular free radical sensors rely on radical coupling reactions[21] and inevitably suffer from signal accumulation over time. These probes are also highly prone to photobleaching, which limits their long-term stability and quantitative accuracy. In contrast, the most distinctive feature of FNDs lies in its robustness as a biosensor that neither depletes over time nor accumulates historical signals, enabling reliable and real-time monitoring of biological processes. In addition, FNDs exhibit exceptional photostability and biocompatibility, making them highly suitable for long-term biological imaging with quantum sensing-based monitoring,[22] as opposed to conventional static measurements or simple trajectory tracking. Previously, FNDs were employed for rotational tracking to assess cellular enzyme dynamics,[23] metabolic activities, and physiological states in living cells under varying conditions.[24]

In recent years, significant advancements have been achieved in the field of paramagnetic species sensing through the application of FND-relaxometry. Early studies were primarily performed in vitro and in buffered environments, demonstrating that NV-based relaxation measurements exhibit exceptional sensitivity toward paramagnetic species such as gadolinium[25] as well as nitroxide radicals[26] and hydroxyl radicals,[21,27] with detection capabilities approaching the single-molecule level.[28] Measurements were conducted on localized chemical events, including pH variations[19] and the oxidation of ascorbic acid.[26] The research has made rapid progress, achieving a substantial leap from principle verification to application in living organisms.[29] Research has successfully extended into the intracellular environment of living cells, and detecting intracellular radicals in-situ and monitoring their real-time changes during incubation with the targeting ligands in a single living cell was achieved.[30] By functionalizing the surface of nanodiamonds, FNDs can be specifically targeted to mitochondria. This capability enables high-specificity detection of metabolic activity and oxidative stress in individual mitochondria, thereby improving spatial precision in intracellular sensing.[31,32] The development of FND-based relaxometry, along with the innovation and integration of various FND technical methods—such as tracking technologies, thermometry, and spectroscopic measurements—has established a robust methodological foundation for studying paramagnetic species in life sciences and materials science. These advances offer promising applications in investigating metabolic dynamics, elucidating how cells orchestrate their internal environment and adapt to changing conditions, and revealing how dysregulation of these processes contributes to diseases such as cancer, metabolic disorders, and neurodegenerative diseases.

In this study, we present a simultaneous integration of spatial position tracking with FNDs relaxometry. The rate of change in movement trajectory provides insight into the activity state of the tracked object and the fluidity characteristics of its surrounding cytoplasmic environment. When integrated with relaxation measurements, this approach offers an additional dimension for characterizing the temporal dynamics of metabolic processes, thereby advancing research on the monitoring of cellular processes over time. To enable real-time nanoscale metabolic tracking in complex biological environments, we present a systematic evaluation of FNDs as a platform for intracellular relaxometry, with a focus on assessing interference from physiological conditions unrelated to metabolism and ensuring a fully reversible—rather than cumulative— response of the paramagnetic signal. Realizing FNDs relaxometry full potential will require synergistic advances in nanomaterial design and quantum measurement techniques. Surface functionalization was employed to direct FNDs to specific organelles, including the nucleus and mitochondria. This approach not only enables subcellular metabolic tracking but also allows for a direct comparison between the internal and external microenvironments within the same living cell nucleus, offering significant potential for elucidating the spatial and functional heterogeneity of cellular metabolism.

## 2. Results and discussion

### 2.1. Evaluation of intracellular FNDs relaxometry

FNDs exhibit stable and bright far-red emission, which is inherently biocompatible and associated with low cytotoxicity. Figure 1a shows a typical image of 40 nm FNDs in a fixed HeLa cell, featuring a confocal image of FNDs (left panel) and a wide-field image displaying merged channels (right panel) for FNDs (red), nucleus (DAPI, blue), and mitochondria (MitoTracker™ Green FM, green). Beyond serving as a fluorescence marker, FNDs can also be applied for intracellular relaxometry.

Relaxometry refers to the measurement of how a quantum system, such as NV centers, returns to thermal equilibrium following a perturbation. The longitudinal relaxation time ($T_1$) represents the time constant associated with the recovery of the population distribution among spin energy levels to their thermal equilibrium state. In the context of FNDs containing NV centers, The $T_1$ relaxation time describes the recovery of the spin system from the polarized $|0\rangle$ state to the thermal equilibrium distribution between the $|m_s = 0\rangle$ and $|m_s = \pm 1\rangle$ ground states. $T_1$ is one of the key characteristic timescales and exhibits high sensitivity to magnetic noise, particularly from paramagnetic species such as free radicals and metal ions. This sensitivity forms the foundation of intracellular relaxometry and various intracellular sensing applications. The electronic structure of the negatively charged NV center (NV$^-$, electron spin S = 1) enables optical initialization and readout of its spin sublevels. A defining characteristic of NV$^-$ is its zero-field magnetic resonance at approximately 2.87 GHz. The relaxation rate ($1/T_1$) is directly proportional to the spectral density of magnetic noise at this specific resonance frequency. Common noise sources, such as lattice vibrations and nuclear spins, typically fluctuate at frequencies far below 2.87 GHz and therefore contribute minimally to the observed relaxation. In contrast, paramagnetic species possess large magnetic moments and exhibit rapid spin fluctuations. Their fluctuation spectra are often broad and include significant components in the gigahertz range. When such a paramagnetic species is in proximity to an NV center, its dynamic spin fluctuations generate magnetic noise at ~2.87 GHz, which can be detected via changes in the NV spin state. This magnetic resonance transition occurs between the $|m_s = 0\rangle$ and $|m_s = \pm1\rangle$ spin states and can be optically probed. Specifically, under continuous illumination with a 532 nm laser, the electron in the NV center is optically pumped into the $|m_s = 0\rangle$ ground state. Once the laser is turned off, the electron gradually relaxes into the $|m_s = \pm1\rangle$ states, eventually reaching thermal equilibrium. In the presence of high-frequency magnetic noise within the intracellular environment, this relaxation process may be significantly shortened. More details of theoretical basis of $T_1$ relaxation measurement is provided in Figure S1.

To achieve intracellular FNDs relaxometry detection, we used a home-built microscope that can perform relaxometry measurements in both wide field and confocal fluorescence microscopy modes under the same setup. This integrated system (Figure 1b) facilitates wide-field $T_1$ measurements via an avalanche photodiode (PD) and confocal $T_1$ measurements using a single-photon avalanche diode (SPAD) while a sCMOS camera enables simultaneous multi-channel imaging under 488 nm and 365 nm excitation, allowing precise organelle localization relative to the FNDs (see Experimental Section for more details). The resolution of this system is comparable to that of a conventional fluorescence microscope, while relaxometry provides localized intracellular information—specifically, magnetic noise within approximately 10 nm around the FNDs.[18] These FNDs act as nanoscale sensors for the local cellular environment. Additionally, the cell culture dish is mounted on the microscope stage under controlled physiological conditions (37 ℃, 5% $CO_2$) to maintain cell viability. HeLa cells cultured under these conditions retained viability exceeding 95% over a three-day period (Figure S2), as confirmed by an automated cell counter.

The relaxation process of a FND within a cell is quantified by monitoring the fluorescent decay. The measurement sequence involves initial spin polarization using a laser pulse, followed by a variable dark time ($\tau$) to allow for spin relaxation, and fluorescence readout via a second laser pulse (Figure 1b inset). The $T_1$ time constant was determined by fitting the observed fluorescence decay. The $T_1$ relaxation dynamics of a single nanodiamond within the fixed HeLa cell presented in Figure 1a was quantified, yielding a value of 148.95 ± 63.23 μs (error bar indicates 95% confidence interval of the fit), as showed in Figure 1c. A histogram depicting the statistical distribution of relaxation times for 32 randomly selected FNDs from a single HeLa cell is presented in Figure 1c insert, with a mean value of 124.11 ± 16.29 μs (mean ± standard deviation). The variation in $T_1$ can be attributed to two primary factors: firstly, the surface and internal properties of the FNDs, and secondly the heterogeneity of the intracellular environment.

The long-term stability of $T_1$ relaxation times, a fundamental requirement for their application in biological systems, was systematically validated across multiple environments. Initially, the stability of FND-based relaxometry was confirmed under biologically relevant solution conditions. In this experiment, FNDs were deposited on a glass-bottom petri dish and their fluorescence was collected using an avalanche photodiode operating in wide-field mode. The measured $T_1$ values thus represent an ensemble average across multiple FNDs. As demonstrated in Figure 1d, FNDs exhibited sustained $T_1$ relaxation stability across aqueous and saline environments—including water, MES buffer (pH 6.0), PBS, and 1M NaCl. Furthermore, nearly identical $T_1$ values were observed in each medium (Figure S3). This robustness to varying chemical conditions underscores the suitability of FNDs for biological measurement contexts where electrolyte composition and pH may fluctuate.

Subsequently, the stability was further verified in fixed-cell systems, which represent a more complex biological environment while eliminating dynamic metabolic variables. In this configuration, fluorescence was collected using a single-photon counter in confocal mode, enabling measurements of individual FNDs. During a continuous one-hour measurement period, the relaxation characteristics of FNDs within fixed HeLa cells exhibited consistent stability. As illustrated in Figure 1e, for 4 particles across 3 fixed HeLa cells, although some numerical fluctuations in $T_1$ values were observed, these variations resulted primarily from measurement signal-to-noise limitations rather than temporal change. Statistical analysis using a linear regression t-test confirmed the absence of a significant monotonic trend ($p > 0.05$), with $T_1$ values distributed relatively evenly around the mean $T_1$ value calculated over the entire measurement period. Measurements of an additional 6 particles across the same 3 fixed HeLa cells (Figure S4) also revealed stable $T_1$ relaxation times. This stability in fixed cells indicates that the nanodiamond sensors are not substantially affected by static cellular components or subcellular structures, supporting their reliability for intracellular applications.

The specificity of the intracellular relaxometry is based on the fact that most of the substances constituting cells and their environment are diamagnetic, and the relaxation properties of FNDs are not affected by diamagnetic substances. However, its response to paramagnetic substances is rapid and reversible. The reversible response of the FNDs to paramagnetic substances was characterized using a wide-field detection scheme with a PD for ensemble-averaged $T_1$ measurements. As evidenced in Figure 1f, the $T_1$ relaxation time in a paramagnetic $GdCl_3$ solution (10 μM) is significantly shorter than that in 100 μM DTPA (diethylenetriaminepentaacetic acid), of which DTPA is a diamagnetic chelate of gadolinium to ensure complete removal of $GdCl_3$. By alternately switching between $GdCl_3$ and DTPA solutions, the $T_1$ relaxation time of FNDs exhibits excellent full reversibility, with no significant signal change observed after 10 cycles (Figure 1g). This demonstrates that the method is reliable for measuring fluctuations in paramagnetic signals. As a non-deplete probe, FNDs does not retain or accumulate previously bound paramagnetic molecules on its surface, thereby preventing potential interference in subsequent measurements. Therefore, FNDs are well suited for real-time and long-term tracking and quantitative measurement over extended ranges.

It is noteworthy that the $T_1$ values in Figure 1d exhibit lower variability, with a coefficient of variation (CV) of 1.3%, compared to those in Figure 1e, which show substantially higher dispersion (CV = 12.3%). This difference arises primarily from the higher fluorescence collection efficiency in wide-field mode, which yields better signal-to-noise ratio within equivalent measurement durations, rather than from environmental factors associated with fixed cells (detailed discussion provided in Supplementary Materials).

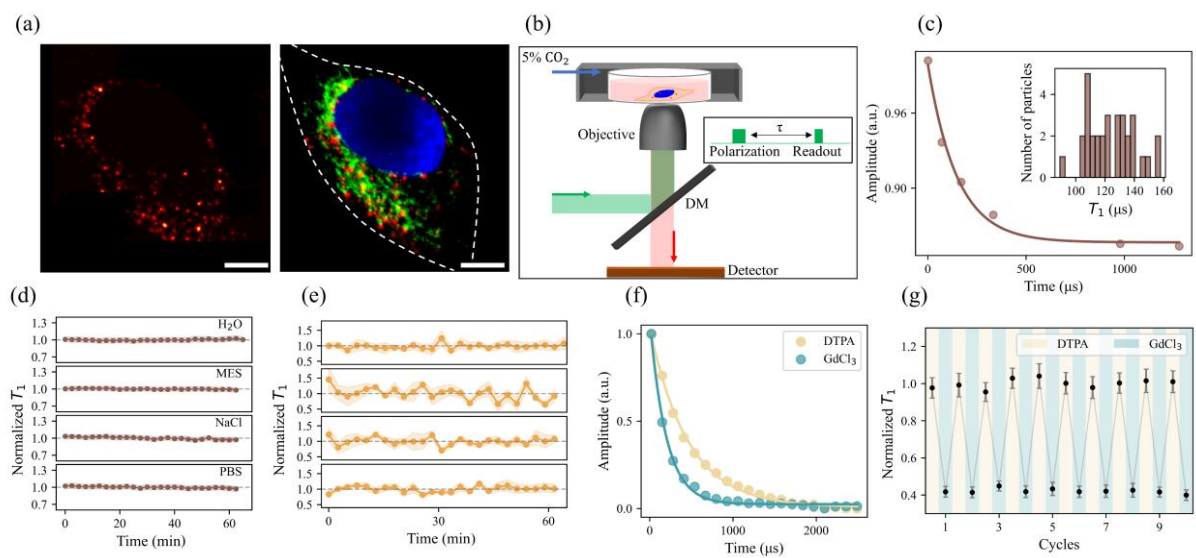

**Figure 1.** Intracellular FNDs relaxation measurement in a fixed HeLa cell and evaluation of their environmental stability and reversibility. (a) Left: fluorescence image of FNDs (red) in a fixed HeLa cell. The image was acquired by a home-built confocal optical setup. Right: Wide-field merged-channel image of the same HeLa cell showing FNDs (red), nucleus (DAPI, blue), and mitochondria (MitoTracker™ Green FM, green). Cell boundaries are indicated by white dotted lines. Scale bars, 10 μm. (b) Schematic diagram of the home-built multimode setup, combing with relaxation measurement, confocal and wide-field imaging. Inset: Timing

sequence for $T_1$ relaxation measurement, comprising an initial laser pulse for spin polarization, a variable dark time ($\tau$) for spin relaxation, and a subsequent laser pulse for fluorescence readout. (c) $T_1$ decay curve of a representative FND from the HeLa cell in (a). The curve was fitted with $I(t) = (Ce^{(-t/T_1)} + 1)I_0$, where C is the fluorescence contrast and $I_0$ denotes the equilibrium fluorescence. The extracted $T_1$ time is 148.95 ± 38.94 μs. Inset: Distribution of $T_1$ times measured from random selected 32 individual FNDs within the same HeLa cell. (d) Time profile of consecutive $T_1$ relaxation measurements for ensemble FNDs in Milli-Q water, MES buffer (pH 6.0), PBS, and 1 M NaCl solution, respectively. (e) Time profile of consecutive $T_1$ relaxation measurements for 4 individual FNDs monitored across three fixed HeLa cells during a 60-minute observation period, demonstrating consistent temporal stability in fixed cells. All $T_1$ relaxation times in (d) and (e) were normalized to the mean $T_1$ value. (f) $T_1$ relaxation curves measured in the presence of the paramagnetic agent $GdCl_3$ (10 μM) (blue curve) and the diamagnetism chelator DTPA (100 μM) (yellow curve). The variation in decay rate reflects the sensing mechanism of FNDs relaxometry in response to changes in the local magnetic environment. (g) Reversible change in $T_1$ values over ten alternating cycles of environmental switching between DTPA (100 μM) and $GdCl_3$ (10 μM), highlighting the reversible nature of the $T_1$ - based sensing mechanism. All $T_1$ relaxation times were normalized to the mean $T_1$ value in DTPA. Error bars in (d) (e) and (g) represent the standard errors of the fits.

## 2.2. FNDs relaxometry in living cell with real-time tracking

Intracellular FNDs relaxometry serves as a method for measuring intracellular magnetic noise and local environmental fluctuations within living cells. In addition, in living cells, FNDs exhibit significantly greater positional dynamics compared to fixed cells, due to both intracellular movement and whole-cell migration. To account for this mobility, we acquired confocal images to map FNDs distributions immediately prior to each individual $T_1$ relaxation time measurement (Figure S5).

The particle tracking methodology operates on the principle that positional drift of a FND within a living cell results in a measurable decrease in fluorescence intensity due to defocusing. To correct for this drift in real-time, a predefined intensity threshold triggers an automated repositioning sequence. Once the signal drops below the threshold, the system initiates a triaxial line-scanning protocol, performing independent linear scans along the x, y, and z axes intersecting at the last known coordinates of FND. Gaussian fitting is then applied to the intensity profiles from each scan, and the peak positions determined from these fits provide the updated, precise three-dimensional coordinates, enabling continuous and robust trajectory reconstruction despite cellular motion or stage drift. This capability is clearly demonstrated by the movement path of a single FND that underwent substantial positional displacement (15.04 μm, the straight-line distance between the initial and final coordinates) within a living HeLa cell, as shown in Figure 2. Figure 2a displays the two-dimensional (XY) trajectory of a single nanodiamond within a living HeLa cell over time, while Figure 2b presents the corresponding three-dimensional trajectory. Figure 2c shows the $T_1$ relaxation times acquired concurrently throughout this tracking process, correlating the spatial position of the FND with its local magnetic environment. The tracking process, underpinned by the drift-correction mechanism, necessarily extends the total measurement time for an equivalent number of $T_1$ sequences compared to controlled environments.

As observed in Figure 2c, the $T_1$ value of the FND exhibited a subtle decrease around 75 minutes and recovered around 120 minutes. Statistical analysis using a linear regression t-test shows a downward trend ($p < 0.05$). To determine whether this fluctuation represented a systematic trend or an isolated occurrence, we subsequently measured $T_1$ relaxation times for six additional FNDs that underwent relatively minor positional drift (3.55 μm in average). Statistical analysis using a linear regression t-test confirmed the absence of a significant monotonic trend ($p > 0.05$) (Figure S6). This comparative analysis indicates that $T_1$ relaxation times generally remain stable in living cells under typical conditions. The minor fluctuation observed in Figure 2c may therefore be attributed to alterations in the local magnetic noise environment resulting from extensive positional displacement, rather than representing a fundamental limitation of the sensing methodology. This finding underscores the importance of considering trajectory-specific environmental factors when interpreting single-particle relaxometry data in dynamic cellular systems. Furthermore, for FNDs localized on organelles, the particle trajectories serves as a tracer for organelle movement.

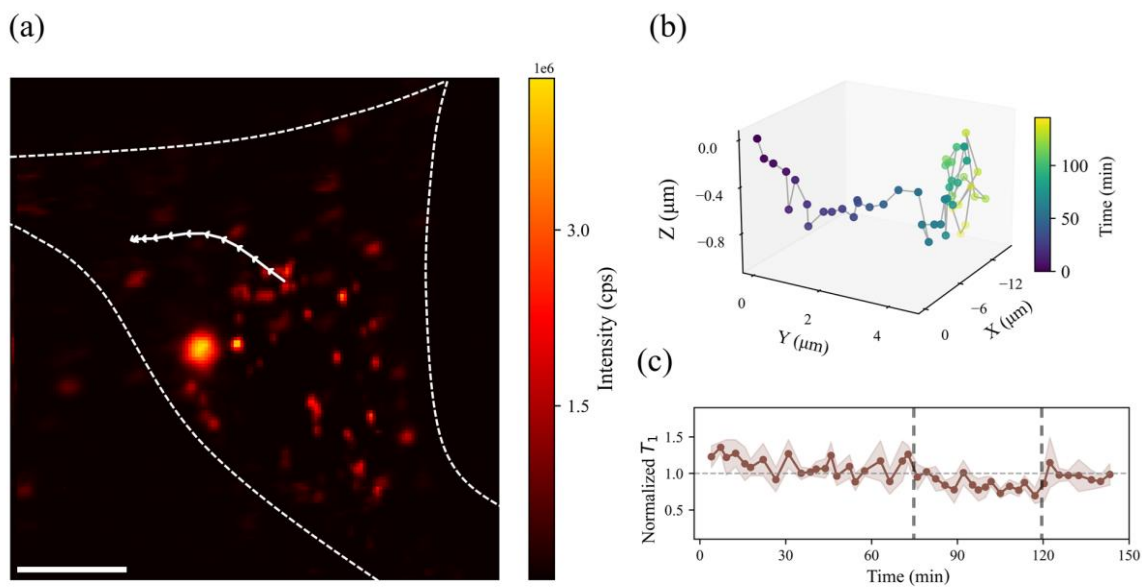

**Figure 2.** Single-FND tracking and $T_1$ dynamics in a living HeLa cell. (a) Confocal microscopy image depicting the intracellular distribution of FNDs in a living HeLa cell. A white trajectory with arrows is XY trajectory of a single FND showing a displacement of 15.04 μm over a period of 140 minutes. White dotted lines mark cell edges. Color bar: Intensity of FNDs. (b) 3D trajectory of the same FND in (a). Color bar: Time period of the whole tracking process. (c) Consecutive $T_1$ relaxation measurements during FND tracking. All $T_1$ relaxation times were normalized to the mean $T_1$ value. Scale bar: 10 μm. Errors bars represent the standard error of the fits.

A key advantage of FNDs lies in their function as solid-state sensors that can be spatially tracked. Therefore, rather than providing a single averaged $T_1$ value, a typical measurement yields a spatially resolved $T_1$ relaxation map across various subcellular compartments. However, to measure the $T_1$ values within different organelles, such as the cytoplasm, mitochondria and nucleus, surface functionalization is necessary to ensure efficient particle targeting to these specific locations. These aspects will be discussed in the following sections.

### 2.3. Subcellular metabolic tracking

Mitochondria are desirable targets for studying cell metabolism because they are the primary sites of energy production and play a central role in many metabolic processes. A markedly shortened $T_1$ relaxation time within mitochondria serves as a direct proxy for both metabolic activity and redox state. MTS (Mitochondria-Targeting Sequence) is a peptide sequence that directs proteins to the mitochondria. It typically consists of a positively charged, hydrophobic amino acid-rich domain, which interacts with the mitochondrial membrane, allowing the protein to be recognized and transported into the mitochondria. MTS sequences are recognized by the mitochondrial import machinery, which facilitates the translocation of the protein across the mitochondrial membranes, ensuring that it reaches the organelle to perform its function. This targeting capability can assist FNDs in effectively localizing to mitochondria. In this study, we employed a synthetic MTS peptide (MLSLRQSIRFFKPATRTLCSSRYLL)[33] to direct FNDs to mitochondria. Successful surface modification was confirmed by DLS analysis (Table 1), which showed a shift in zeta potential from negative (-12.86 ±1.45 mV) to positive (25.37 ±2.23 mV) and an increase in hydrodynamic diameter.

Imaging performed with a confocal microscope revealed minimal colocalization between bare FNDs (red) and mitochondria (green) (Figure 3a), whereas significant colocalization was observed for MTS-conjugated FNDs (FND-MTS; red) with mitochondria (green) (Figure 3b). Manders coefficients[34] were used to quantify the degree of colocalization between two different fluorescent signals between FNDs and mitochondria in fluorescence microscope images. These coefficients are calculated based on the overlap of fluorescence intensities from two distinct fluorophores in a given region of interest. To quantify targeting efficiency, Manders coefficients ($M_1$ and $M_2$) were calculated as follows: after applying intensity thresholds for each channel, pixels below the threshold were set as zero. $M_1$ represents the fraction of signal in channel 1 (FNDs) that colocalizes with channel 2 (mitochondria), and $M_2$ represents the fraction of mitochondria colocalizing with FNDs. These are defined by:

$$M_1 = \frac{\sum_i I_{coloc,i}}{\sum_i I_i} \quad (1)$$

$$M_2 = \frac{\sum_j I_{coloc,j}}{\sum_j I_j} \quad (2)$$

$$I_{coloc,i} = \begin{cases} I_i & \text{if } I_j > 0 \\ 0 & \text{otherwise} \end{cases} \quad (3)$$

$$I_{coloc,j} = \begin{cases} I_j & \text{if } I_i > 0 \\ 0 & \text{otherwise} \end{cases} \quad (4)$$

Here, i and j denote pixel in the FND and mitochondrial channels, respectively. The subscript coloc refers to the subset of thresholded pixels that also exhibit colocalization.

Since not all mitochondria are expected to colocalize with FNDs, $M_1$ was used to evaluate targeting efficiency. The $M_1$ value for FND-MTS (0.60 ±0.07) was significantly higher than that of bare FND (0.16 ±0.06) (Figure 3c). Besides, FND-MTS exhibit pronounced aggregation specifically localized in proximity to mitochondria, while maintaining a dispersed distribution in other cytoplasmic regions (Figure S7). In contrast, bare FNDs show no such specific aggregation and are uniformly distributed throughout the cell (Figure S8). The difference of spatial distribution provides evidence that the MTS peptide modification effectively directs the FNDs to mitochondria, leading to localized accumulation. Both bare FND and FND-MTS were applied at a concentration of 20 μg mL$^{-1}$ and incubated with HeLa cells for 24 h. We assessed the viability of HeLa cells following a 24-hour incubation with bare FNDs and mitochondrial-targeted FNDs (FND-MTS) at various concentrations. Both bare FND and FND-MTS exhibited excellent biocompatibility at low concentrations (Figure S9).

Nonetheless, in an image with conventional resolution, it remains arduous to differentiate whether a FNDs is localized within mitochondria or in the cytoplasm adjacent to mitochondria. The environmental sensing capacity of nanoparticles is predominantly within a range of 10 nanometers, which is far smaller than the range of fluorescence resolution. Consequently, accurately quantifying the relaxation times of FNDs at mitochondrial and cytoplasmic sites for comparison poses a significant challenge. However, upon stimulation with specific drugs, the relaxation alterations induced by mitochondria far exceed those at the locations of other organelles. Carbonyl cyanide m-chlorophenyl hydrazone (CCCP), an uncoupler of oxidative phosphorylation, was used to induce mitochondrial stress and increase ROS levels. Upon addition of 20 μM CCCP, the $T_1$ relaxation time of a single nanodiamond in a living HeLa cell decreased significantly (Figure 3d). To unequivocally attribute this $T_1$ reduction to the biological effect of CCCP rather than any non-specific chemical interaction between CCCP and the surface of FNDs, a critical control experiment was performed. We measured the $T_1$ relaxation times of 8 individual FND-MTS exposed to an identical concentration of CCCP (20 μM) in a cell-free environment. No significant change in $T_1$ was observed in any of these control measurements (Figure S10), confirming that the pronounced decrease in $T_1$ recorded within cells is a specific consequence of the drug-induced oxidative stress. This finding, when considered alongside the stable $T_1$ values observed under basal conditions, provides critical validation of reliability of $T_1$ relaxometry. The combination of baseline stability and stimulus-evoked response demonstrates that the sensors are neither passivated nor impaired by the intracellular environment. Instead, they function as dynamic reporters of redox status.

**Table 1.** Dynamic Light Scattering results.

| Sample | Size [nm] | Zeta potential [mV] | Polydispersity index |
| --- | --- | --- | --- |
| Bare FND | 57.99 ±0.96 | -12.86 ±1.45 | 0.153 ±0.013 |
| FND-TAT-NLS | 81.85 ±1.05 | 35.03 ±0.40 | 0.127 ±0.004 |
| FND-MTS | 76.96 ±0.41 | 25.37 ±2.23 | 0.160 ±0.008 |

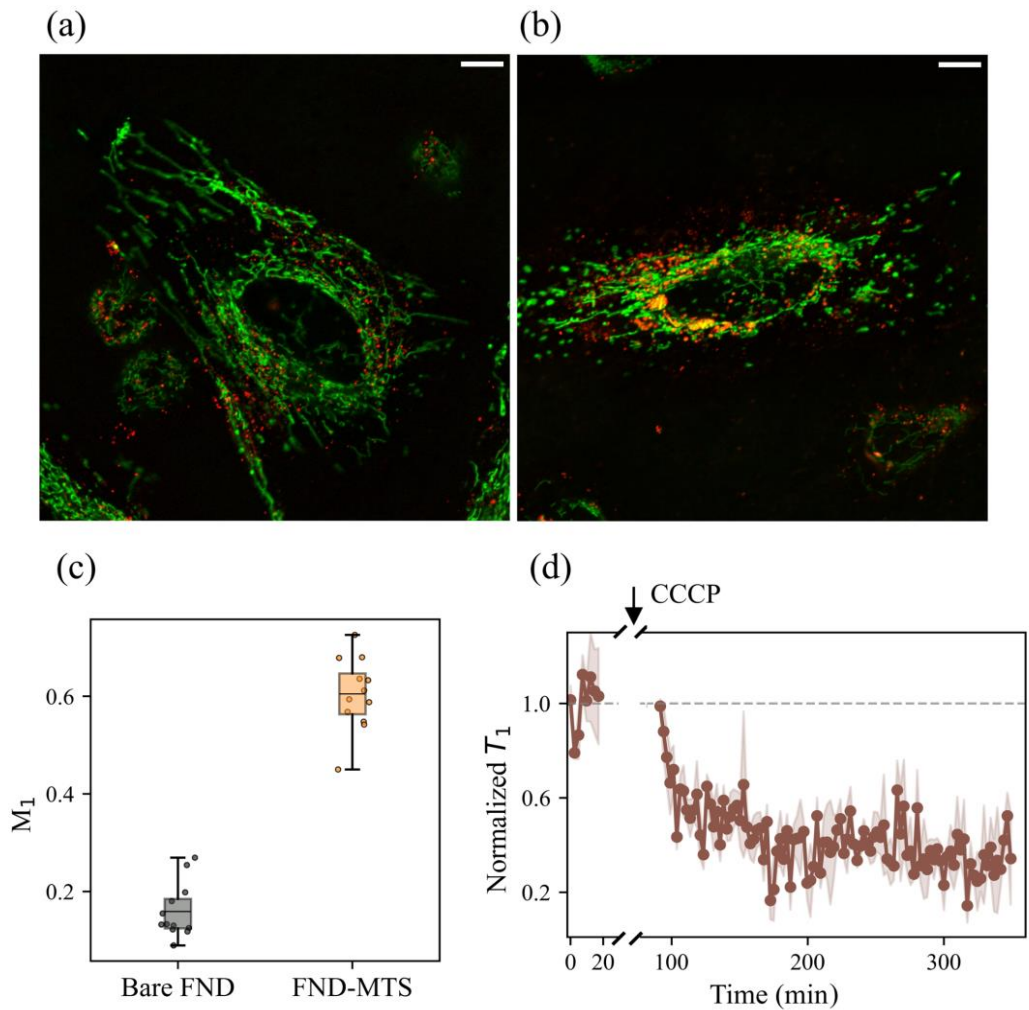

**Figure 3.** Mitochondrial targeting of FND-MTS and monitoring of oxidative stress via $T_1$ relaxometry. (a) Confocal image of HeLa cells after 24-hour incubation with bare FNDs (20 μg mL$^{-1}$), demonstrating minimal colocalization between FNDs (red) and mitochondria (MitoTracker™ Green FM, green). (b) Confocal image of HeLa cells after 24-hour incubation with FND-MTS, showing significant colocalization (red) with mitochondria (green), confirming mitochondrial targeting. Scale bars: 10 μm. (c) Quantitative analysis using Manders' coefficient ($M_1$), revealing significantly higher mitochondrial targeting efficiency for FND-

MTS compared to bare FNDs. The box plot depicts the distribution of FND-MTS data, with the center line indicating the average, the box representing the interquartile range, and the bar showing the full data range. (d) Time-dependent decrease in $T_1$ relaxation time following addition of 20 µM CCCP, indicating the potential of FND-MTS as a sensor for oxidative stress. All $T_1$ values were normalized relative to the average $T_1$ value measured before the application of CCCP. Errors bars represent the standard error of the fits.

Targeting the nucleus with nanoparticles is always a challenging task. Because first nanoparticles are often recognized by cells as foreign substances and are encapsulated and excreted by lysosomes. Secondly, as the most important organelle within a cell, the nuclear membrane protects the nuclear region, making it often difficult for nanoprobes to reach the interior. Even if there are probes that can enter the nucleus by chance, their quantity is still very low compared to the total number taken into the cell. Nuclear localization signal (NLS) to interact with the nuclear pore complex can help to target into nucleus. Meanwhile, the long-term stability of NLS-modified particles remains suboptimal. An increasing number of particles settle in the culture medium and accumulate at the bottom of the culture dish, rather than being efficiently loaded by the cells. As viruses usually utilize different peptides to cross each cell membrane barrier, TAT is a generic membrane translocation peptide.[35] When it is combined with NLS and FNDs, it can assist in achieving effective cell membrane translocation. By adjusting the combination mode and proportion of different nuclear localization signals and cell membrane translocation peptides, a relatively higher nuclear localization efficiency was obtained with FND-TAT-NLS. A functionalized nanodiamond (FND-TAT-NLS) was designed by conjugating TAT peptide (GRKKRRQRR) and NLS (PPQPKKKRKV)[36] to the FND surface via electrostatic adsorption. Given the negatively charged surface of carboxyl-terminated FNDs, the positively charged peptides readily adsorb, enabling efficient functionalization. Dynamic light scattering (DLS) analysis (Table 1) confirmed successful surface modification, as evidenced by the shift in zeta potential from negative (-12.86 ± 1.45 mV) to positive (35.03 ± 0.40 mV). The slight increase in hydrodynamic diameter suggests that FND-TAT-NLS retains the size-dependent nuclear entry capability.

To differentiate between nuclear-localized FNDs and those merely adjacent to the nucleus—which may appear colocalized in two-dimensional projections—we acquired high-resolution z-stack images by confocal microscope. Figure 4a demonstrates the absence of colocalization between bare FNDs and the nucleus (Red: FND; Blue: nucleus). In contrast, a subset of FND-TAT-NLS particles exhibited nuclear colocalization, as shown in Figure 4b. A three-dimensional projection of the region circled in Figure 4b (Figure 4c) confirms that the nanodiamond indicated by a white arrow is located within the nucleus, rather than adjacent to it. Furthermore, transmission electron microscopy (TEM) provides high-resolution confirmation of successful FNDs localization within the nucleus of a HeLa cell (Figure 4d). The main panel presents a comprehensive overview of the nuclear compartment. The magnified view in the lower-left inset corresponds directly to the area demarcated by the green square in the main panel. This high-magnification image clearly reveals the distinct signature of FNDs, providing unambiguous verification of successful nuclear entry beyond the resolution limits of conventional optical microscopy. Figure S11 provides additional images of intranuclear FNDs. Both bare FND and FND-TAT-NLS were applied at a concentration of 20 μg mL$^{-1}$ and incubated with HeLa cells for 24 h. We assessed the viability of HeLa cells following a 24-hour incubation with bare FNDs and nucleus-targeted FNDs (FND-TAT-NLS) at various concentrations (Figure S9). It should be noted that the nuclear targeting efficiency of FND-TAT-NLS remains relatively low, which can be attributed to three primary factors. While the central transport channel of human nuclear pores has a diameter of approximately 40 nm,[37] the comparatively large hydrodynamic size of FND-TAT-NLS limits its efficient translocation to only a fraction of the pores. Second, FNDs exhibit a tendency to aggregate in saline environments such as cell culture medium, which may further impede nuclear entry. Third, a fraction of internalized FNDs is likely sequestered within lysosomes or other vesicles, diverting them from the intended nuclear localization pathway. Collectively, these factors contribute to the observed limitations in targeting efficiency. However, in practical measurements, the nucleus can be clearly identified using either bright-field imaging of the cell or confocal scanning imaging of FNDs. Non-nuclear-targeted FNDs tend to distribute around the nucleus, effectively outlining its boundary. Therefore, targeting efficiency poses little impediment to the identification of FNDs located within the nucleus.

$T_1$ relaxation times of FNDs located in the nucleus and cytoplasm were compared to evaluate the influence of different biochemical compositions within cells. Measurements were performed on 43 FNDs in the nucleus and 43 in the cytoplasm across 11 living HeLa cells. As summarized in Figure 4e, no significant difference was observed between the $T_1$ relaxation times in the nucleus (110.65 ± 14.74 μs, mean ± standard deviation) and cytoplasm (114.34 ± 12.53 μs). Although this result did not reveal a statistically significant difference, it provides valuable insight into the basal magnetic noise landscape of the nucleus. The comparable $T_1$ times suggest that under steady-state conditions, the overall magnetic noise environment—arising from paramagnetic species such as free radicals and metal ions—may be similar in both compartments. This offers a new perspective on nuclear redox homeostasis.

Notably, $T_1$ relaxation times are much higher in complete media (228.22 ± 43.15 μs) than those in living HeLa cells. The cytoplasm contains a high concentration of paramagnetic species, including reactive oxygen species and metal ions such as ferritin. These molecules generate strong local magnetic noise that accelerates NV spin relaxation. Besides, the dense cellular environment, rich in proteins and organelles, increases local electronic spin density, further enhancing magnetic noise and thereby significantly reduces the $T_1$ relaxation time. Furthermore, as detailed in Figure 4f, $T_1$ relaxation times showed no considerable variation across the 11 individual HeLa cells examined.

Spatially resolved relaxometry within the nucleus paves the way for novel research directions. The nucleus is central to the regulation of gene expression, DNA repair, and the integration of metabolic signals. It works in close coordination with other cellular structures, such as the mitochondria, to regulate metabolic processes and cellular responses. Studying the nucleus in the context of metabolism helps to understand how the cell coordinates its internal environment and adapts to changing conditions, and how disruptions in these processes can lead to diseases like cancer, metabolic disorders, and neurodegenerative diseases.

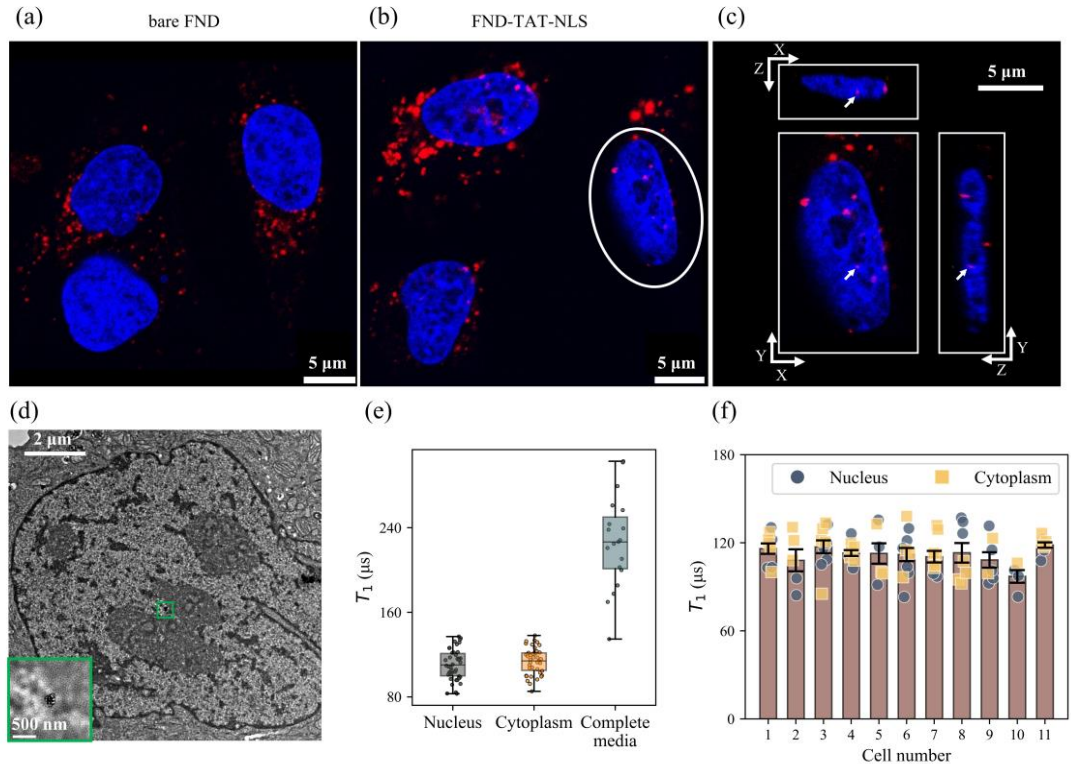

**Figure 4.** Nuclear localization of FND-TAT-NLS and measurement of $T_1$ relaxation times inside nucleus. (a) Confocal image of HeLa cells following a 24-hour incubation with bare FNDs (20 μg mL$^{-1}$), showing an absence of colocalization between FNDs (red) and the nucleus (DAPI, blue). (b) Confocal image of HeLa cells incubated with FND-TAT-NLS (20 μg mL$^{-1}$, 24 h), where a subset of functionalized FNDs (red) were colocalized with the nucleus (blue). (c) Three-dimensional projection of the region circled in (b), confirming the intranuclear localization of the FND indicated by the white arrow. (d) Transmission electron micrograph confirming intranuclear localization of FNDs. (e) Comparison of $T_1$ relaxation times among FND-TAT-NLS in different locations: 43 in the nucleus, 43 in the cytoplasm, and 20 in the complete media. The results indicate no significant difference between the nucleus and cytoplasm, but a significant difference was observed between the intracellular (nucleus/cytoplasm) and extracellular environments. The box plot depicts the distribution of FND-TAT-NLS data, with the center line indicating the average, the box representing the interquartile range, and the bar showing the full data range. (f) $T_1$ relaxation times measured across the 11 individual HeLa cells analyzed, demonstrating consistent values with no appreciable cell-to-cell variation. Data are represented as mean ± standard deviation with individual data points.

## 3. Conclusion and discussion:

In this study, we systematically evaluated FNDs as a probe for intracellular relaxometry in living cells. The assessment focused on FND stability in complex biological environments and the reversibility of paramagnetic signal detection, both of which are critical for enabling dynamic and non-invasive monitoring of intracellular processes. The evaluation encompasses two key components: foundational reversibility, involving the explicit quantification of $T_1$ signal reversibility under controlled chemical conditions; and stability assessment, which entails separate evaluations in chemical environments, fixed cells, and living cells. To accurately capture intracellular dynamics, a single-particle tracking protocol was developed. This multi-tiered analytical framework provides a comprehensive understanding of the technology's robustness across diverse experimental conditions. These characteristics support real-time metabolic probing and facilitate the observation of transient biochemical events. The results demonstrate that FNDs hold significant potential as reliable and responsive tools for investigating cellular signaling mechanisms.

We demonstrate a simultaneous integration of spatial position tracking with FND-based relaxometry, enabling successful measurement of microenvironments within living cells, including mitochondria, the cell nucleus, and the cytoplasm. The rate of change in movement trajectory provides valuable insight into the activity state of the tracked object and the rheological properties of its surrounding cytoplasmic environment. When combined with FND-based relaxometry, this integrated approach adds a distinct dimension for characterizing the temporal dynamics of metabolic processes, thereby enhancing the capability to monitor cellular processes over time. Meanwhile, this study achieves, for the first time, relaxation measurements of the intranuclear microenvironment, enabling direct comparative analysis between the internal and external environments within the same living cell nucleus. This approach is crucial for investigating the spatial and functional heterogeneity of cellular metabolism.

Despite this, there are still many limitations in using FNDs relaxometry to study intracellular metabolism. Firstly, the properties of FNDs, mainly refers to the $T_1$ relaxation time in this work, exhibit inherent heterogeneity,[3,38] and the resulting variations in characteristics across individual particles pose significant challenges for standardization and reproducibility. We analyzed the $T_1$ relaxation properties of 87 individual FNDs under dry conditions (Figure S12). These results reveal that intrinsic heterogeneity in particle characteristics persists even in the absence of the complex intracellular biochemical environment. Secondly, the targeting efficiency of the particles in living cells remains suboptimal, limiting their reliable delivery to specific intracellular sites. The size of FNDs can influence their capacity to achieve precise intracellular localization with high targeting efficiency. The 40-nm-sized particles used in this work were selected after weighing various sizes and properties. Starting from the goal of achieving intracellular localization, FNDs of a size similar to that of GFP (5nm) are applicable. However, most FNDs of such a size do not possess a crystal structure and thus cannot be used in quantum sensing applications. In this regard, further improvement of the properties of small particles is necessary. Currently, particles with excellent relaxation properties can be prepared by the chemical vapor deposition (CVD) method.[8,39] However, how to obtain high-quality particles in large quantities remains an important issue. Thirdly, compared with some commonly used fluorescent dyes, such as rhodamine, or fluorescent nanoparticles, such as quantum dots, small FNDs are not very bright. In the condition with strong autofluorescence or interference from other dyes, time-gated detection can effectively suppress background signals, significantly enhancing imaging contrast.[40] All-optical lock-in imaging methods leveraging NV spins have demonstrated high-contrast visualization of nanodiamond-labeled biomolecules, with signal-to-background ratios enhanced by up to two orders of magnitude relative to conventional fluorescence imaging techniques.[41] Last but not least, the factors influencing intracellular FNDs relaxometry extend far beyond free radicals, which encompass a wide variety of distinct species. Specificity and selectivity continue to represent a major challenge for virtually all probe platforms. Spectroscopic resolution is an advantage in NV center-based sensing. Looking ahead, significant advances in FNDs for nanoscale nuclear magnetic resonance (NMR)[8] and electron spin resonance (ESR)[9] sensing offer immense potential for in vivo measurements—such as single-cell-level ESR and NMR—to enable in-depth analysis and a profound understanding of metabolic activities underlying biological processes.

Overall, our findings provide a solid foundation for the future development of advanced metabolic imaging technologies with subcellular resolution. The integration of FNDs into cellular imaging workflows represents a promising advancement toward real-time monitoring of metabolic processes.

**Experimental Section**

*Applied materials*: FNDs were purchased from Adamas Nanotechnologies (NDNV40nmHi). Phosphate-buffered saline (PBS, 1×), minimum essential medium (MEM) and HeLa cells (human cervical cancer cells) were obtained from Sunncell. MES buffer (pH =6.0) were obtained from Maokangbio. Sodium chloride, ethanol and acetone were obtained from Sinopharm Chemical Reagent Co., Ltd. (SCRC). 4% formaldehyde fixative (in PBS, methanol free), 0.4% trypan blue solution, fetal bovine serum (FBS), trypsin, penicillin/streptomycin, lead citrate, MitoTracker™ Green FM and 4′,6-diamidino-2-phenylindole (DAPI) fluorescent dyes were obtained from Thermo Fisher Scientific. Gadolinium (III) chloride ($GdCl_3$) was obtained from Macklin. Diethylenetriaminepentaacetic acid (DTPA) was obtained from MedChemExpress. TAT-NLS peptide and MTS peptide were synthesized according to published procedures. Cell Counting Kit-8 was obtained from Biosharp. 96-well plate was obtained from Corning. Glutaraldehyde, osmium tetroxide and cacodylate buffer were purchased from Electron Microscopy Sciences. Paraformaldehyde was purchased from Ted Pella, Inc. Potassium ferrocyanide was purchased from Sigma Aldrich. Uranyl acetate and epoxy resin were purchased from SPI Supplies.

*Optical setup for $T_1$ relaxation time measurement*: A home-built dual-modality microscopy platform with three excitation wavelengths was constructed for confocal and wide-field measurements (Figure M1). A continuous-wave 532nm laser (Verdi G Series 5W, Coherent) was used for NV excitation, a 488nm laser (MDL-III-488, CNIlaser) was used for MitoTracker™ Green FM excitation and a 365nm LED (LEM365C1, JCOPTIX) was used for DAPI excitation. An acousto-optic modulator (M1133-aQ80L-3, ISOMET) was used to control the on-off of 532nm laser. A 500 MHz PulseBlaster (PulseBlasterESR-Pro, Spincore Technologies) was used to control the pulse sequence. A microscope objective (LUCPLFLN60X, N.A. 0.7, Olympus) was used for excitation and fluorescence collection. A 550nm long pass dichroic beam splitter (DMLP550L, Thorlabs) was used for reflection of 532nm laser and transmission of the fluorescence of FNDs. A 425nm long pass dichroic beam splitter (DM20-425LP, LBTEK) was used for reflection of 365nm light and transmission of the fluorescence of DAPI. A 505nm long pass dichroic beam splitter (DMLP505L, Thorlabs) was used for reflection of 488nm light and transmission of the fluorescence of MitoTracker™ Green FM. The dichroic beam splitters were exchanged with the magnetic retention bases (MKB-75, LBTEK). For wide-field detection, a lens (AC508-400-A-ML, Thorlabs) was used to focus the 532nm laser on the back focal plane of the objective to illuminate the entire area of the field of view. To focus 488nm laser, a lens (MAD424-A, LBTEK) was used. The 365nm light was directly reflected to the back focal plane without extra focusing. The fluorescence was detected with a sCMOS camara (400BSI, Dhyana) or an avalanche photodiode (APD410A/M, Thorlabs). Two flip mirrors were used to change the detector module. For confocal detection, only 532nm laser was used, the laser was focused with the objective and the fluorescence was detected with a single-photon avalanche diode (SPCM-AQRH, Excelitas). To assure cell viability, a live-cell incubator (STX TIZWX, Tokaihit) was fixed on the sample stage.

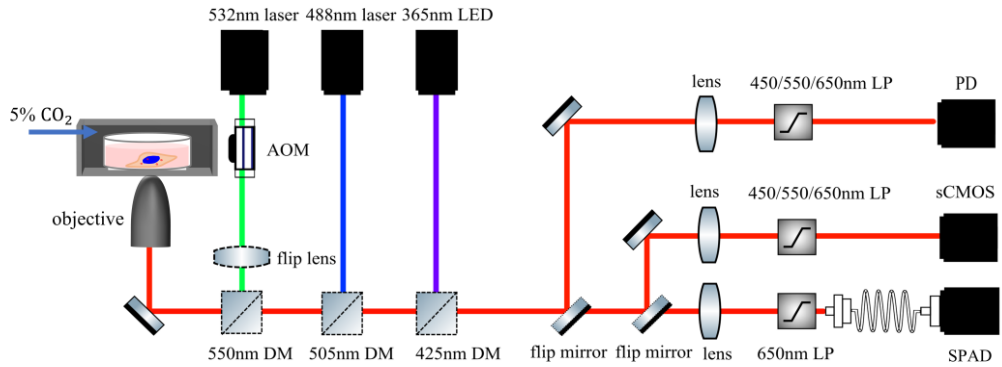

Figure M1. Experimental setup for $T_1$ relaxation time measurement. AOM: acousto-optic modulator. DM: dichroic beam splitter. LP: long pass filter. PD: avalanche photodiode. SPAD: single-photon avalanche diode.

*Measurement sequence of $T_1$ relaxation time and data processing*: For confocal measurement, the NV centers in the FNDs were optically polarized into the $|m_s = 0\rangle$ ground state using a 10 μs, 532 nm laser pulse. Following a variable waiting time τ, a second laser pulse of 1.4 μs was applied to read out the spin state. The SPAD was gated on at 800 ns after the onset of the readout pulse, collecting fluorescence for 500 ns to obtain the signal intensity. Subsequently, a 15 μs laser pulse reinitialized the NV center. After a 10 μs delay, an identical 1.4 μs readout pulse was applied, with the APD again gated from 800 ns to 1300 ns to acquire the reference intensity. The normalized signal was calculated as the ratio of the signal to the reference intensity. This measurement sequence was repeated tens of thousands of times to achieve sufficient signal-to-noise ratio. For wide field measurement, the NV centers in the FNDs were optically polarized using a 15 μs, 532 nm laser pulse. Following a variable waiting time τ, a second laser pulse of 11 μs was applied to read out the spin state. The PD was gated on at 1 μs after the onset of the readout pulse, collecting fluorescence for 10 μs to obtain the signal intensity. After a 10.033 ms delay, an additional readout pulse was applied to measure the fluorescence at the equilibrium state. Subsequently, a 1 μs, 532 nm laser pulse was applied, followed by another readout pulse to measure the fluorescence at the polarized state. The contrast was calculated as the difference between the fluorescence intensity at the polarized state and that at the equilibrium state. The initially collected signal was then normalized by this contrast. This sequence was also repeated tens of thousands of times to ensure an adequate signal-to-noise ratio.

For data analysis, the averaged normalized signal was fitted to the following equation to extract the $T_1$ time constant:

$$I(t) = (Ce^{(-t/T_1)} + 1)I_0 \tag{5}$$

Where C is the fluorescence contrast and $I_0$ denotes the equilibrium fluorescence intensity. The 95% confidence intervals for the fitted parameters were calculated as the estimated value plus/minus the standard error multiplied by the critical value from the t-distribution with (n-p) degrees of freedom.

The detection of linear regression trends through p-value analysis follows a rigorous statistical protocol. A simple linear regression model is first fitted to the data using least-squares estimation, establishing the mathematical relationship between variables. The core analysis focuses on testing the null hypothesis that the slope coefficient equals zero, indicating no linear relationship, against the alternative hypothesis of a non-zero slope. The statistical significance is determined through a t-test comparing the estimated slope to its standard error, generating a p-value from the t-distribution with appropriate degrees of freedom. This p-value provides the quantitative basis for decision-making: values below the predetermined significance threshold (typically $\alpha = 0.05$) reject the null hypothesis, supporting a statistically significant linear trend, while values above this threshold indicate insufficient evidence for such a relationship.

*Surface functionalization of FNDs*: FND-TAT-NLS: Raw FNDs (250 μg) were dispersed in 250 μL deionized water and sonicated for 15 min. Subsequently, excess TAT-NLS peptide (2 mg in 250 μL deionized water) was added to the FNDs suspension, and the mixture was stirred for 12 h at room temperature and stored at 4 ℃. The suspension was subjected to an additional 15 minutes of sonication prior to use to ensure adequate monodispersity. FND-MTS: Raw FNDs (250 μg) were dispersed in 250 μL of deionized water and sonicated for 15 minutes. To solubilize the mitochondrial targeting signal (MTS) peptide, 62.5 μL of dimethyl sulfoxide (DMSO) was mixed with 187.5 μL of deionized water, after which 3 mg of MTS peptide was dissolved in the DMSO–water mixture. MTS solution was then added to the FNDs suspension, and the mixture was stirred for 12 hours at room temperature and stored at 4 ℃. The suspension was further sonicated for an additional 15 minutes prior to use to ensure optimal monodispersity.

*Characterization of FNDs*: Particle size (hydrodynamic diameter) distribution and zeta potential were characterized via dynamic light scattering (DLS) using a NanoBrook Omni system (Brookhaven Instruments). The hydrodynamic diameter was measured by diluting synthesized FNDs in Milli-Q to ~0.01 mg mL$^{-1}$ with a 90 °angle at room temperature. The zeta potential was measured at room temperature. All experiments were performed in triplicate, with data presented as mean ± standard deviation.

*Cell cultivation and FND loading*: HeLa cells were cultured in MEM supplemented with 10% FBS and 1% penicillin/streptomycin, under a humidified atmosphere of 5% $CO_2$ at 37 °C. When the cells reached approximately 95% confluence, they were detached using a standard trypsinization protocol and subsequently reseeded into 35 mm glass-bottom petri dishes (Casmart). The cells were then incubated in culture medium until 70% confluence was achieved. Subsequently, different samples at specified concentrations were added to the cells, followed by an additional 24-hour incubation period prior to further characterization.

*Cell staining*: Nuclear staining. HeLa cells were reseeded into 35 mm glass-bottom petri dishes, washed twice with PBS, and fixed with 4% formaldehyde for 15 minutes. After fixation, cells were rinsed three times with PBS. The cells were then incubated with a 300 nM DAPI solution in PBS for 5 minutes, followed by three additional PBS washes. Mitochondrial staining. HeLa cells were reseeded into 35 mm glass-bottom Petri dishes, and no fixation step was required. The cells were directly incubated with a 20 nM MitoTracker™ Green FM solution in complete medium for 45 minutes, followed by three washes with PBS.

*Target efficiency assessment*: The targeting efficiency of FND-MTS was assessed using two confocal microscopy setups. Mitochondrial targeting was evaluated with a Zeiss LSM880 microscope equipped with a Plan-Apochromat 63×/1.4 Oil DIC M27 objective. FNDs were excited at 543 nm and detected at 585–733 nm (pinhole: 0.99 AU; detector gain: 761.4 V), while mitochondria were labeled with MitoTracker™ Green FM (excitation: 488 nm; detection: 493–600 nm; pinhole: 1.73 AU; detector gain: 857.1 V). Each image was acquired with 8-frame averaging. Nuclear targeting was evaluated using a Zeiss LSM980 microscope with the same objective model. FNDs were excited at 590 nm and detected at 595–757 nm (pinhole: 1.00 AU; detector gain: 663 V), and nuclei were stained with DAPI (excitation: 353 nm; detection: 410–581 nm; pinhole: 1.00 AU; detector gain: 593 V). Each image was acquired with 4-frame averaging. To accurately distinguish FNDs within the nucleus from those above or below it—which may appear colocalized in the xy-plane—z-stacks were collected at 0.2 μm intervals. Three-dimensional colocalization analysis was performed using orthogonal projections generated in ZEN Lite 3.9 software (Zeiss), while two-dimensional mitochondrial colocalization was quantified using the JACoP plugin in ImageJ. Thresholds for both channels were set automatically via Costes Automatic Thresholding Approach within JACoP. Manders coefficients were calculated from 12 images (each containing 1–4 cells), and results are presented as the average Manders' coefficient across all images. For all images, the display window of the channel of FNDs was adjusted by linearly stretching the grayscale range to ensure a 0.16% pixel saturation level.

*Biocompatibility assessment of FND, FND-TAT-NLS and FND-MTS using CCK-8 assay:* The biocompatibility of FND, FND-TAT-NLS and FND-MTS was evaluated using the Cell Counting Kit-8 (CCK-8) assay. HeLa cells were seeded in 96-well plates at a density of $2 \times 10^3$ cells per well in 100 μL complete growth medium and allowed to adhere overnight. After cell adhesion, FND, FND-TAT-NLS and FND-MTS were added to each well at concentrations ranging from 0 to 50 μg mL$^{-1}$ (0, 0.5, 5, 10, 20, 50 μg mL$^{-1}$). Two control groups were established: (1) negative control (untreated cells in complete medium) and (2) blank control (cell-free wells containing medium only). After 24 h incubation, 10 μL of CCK-8 solution was added to each well, followed by additional incubation for 1.5 h at 37 ℃. Absorbance was measured at 450 nm using a microplate reader (SpectraMax iD5, Molecular Devices). Cell viability was calculated according to the following equation:

$$\text{Viability}(\%) = \frac{\text{OD}_{\text{sample}} - \text{OD}_{\text{blank}}}{\text{OD}_{\text{control}} - \text{OD}_{\text{blank}}} \times 100\% \tag{6}$$

All experiments were performed in triplicate, with data presented as mean ± standard deviation.

*Cell viability test of the stage-top incubator*: HeLa cells were seeded in 35mm petri dish and incubated with the stage-top incubator. For viability assessment, cell suspensions were diluted to appropriate concentrations in MEM, then mixed 1:1 (v:v) with 0.4% trypan blue solution. Viable cell counts were determined using an automated cell counter (Invitrogen Countess 3 FL, Thermo Fisher Scientific) following manufacturer protocols. Triplicate measurements were performed for each experimental condition, with viability calculated as the percentage of trypan blue-excluding cells relative to total cell count. Data were presented as mean ± standard deviation.

*Cell Sample Preparation for Electron Microscopy:* Cell samples were prepared through a comprehensive processing protocol beginning with primary fixation using 3% glutaraldehyde and 2% paraformaldehyde in 0.1 M phosphate buffer (pH 7.2) for 2 hours at room temperature. Following initial fixation, samples underwent secondary fixation through sequential treatments with potassium ferrocyanide and osmium tetroxide in cacodylate buffer, followed by en bloc staining with uranyl acetate under light-protected conditions. Dehydration was then performed through a graded ethanol series followed by acetone rinses, ensuring complete water removal while preventing air exposure artifacts. Samples were subsequently infiltrated with epoxy resin using progressively increasing resin concentrations in acetone mixtures, culminating in pure resin infiltration. The polymerization was achieved through a two-stage thermal curing process at 45 ℃ for 12 hours followed by 60 ℃ for 24 hours. Following resin polymerization, ultrathin sections were prepared at a thickness of 70 nm using a Leica UC7 ultramicrotome. Sections were collected on formvar-coated copper grids and post-stained with uranyl acetate and lead citrate before TEM imaging.

*Transmission electron microscopy analysis:* Transmission electron microscopy was performed using a Tecnai T12 Biotwin microscope (FEI Company) operating at an acceleration voltage of 120 kV. Sample analysis employed a specialized defocus-based identification protocol to distinguish FNDs from potential osmium tetroxide precipitation artifacts. Following localization of electron-dense particles, controlled defocusing was implemented to exploit the crystalline nature of FNDs. The characteristic intensity modulation observed under defocused conditions—resulting from Bragg diffraction contrast mechanisms—provided unambiguous identification of diamond crystallites, while amorphous osmium deposits remained visually static under identical defocus conditions.

*Assessment of $T_1$ relaxation time of FNDs in complete media*: A suspension of FNDs in deionized water (40 μL, 0.1 mg mL$^{-1}$) was spin-coated onto a petri dish at 6000 rpm for 90 seconds. $T_1$ relaxation measurements were performed using confocal mode after adding 2 mL of complete media.

*Assessment of $T_1$ relaxation time of FND-MTS in CCCP*: A suspension of FND-MTS in deionized water (40 μL, 0.1 mg mL$^{-1}$) was spin-coated onto a petri dish at 6000 rpm for 90 seconds. $T_1$ relaxation measurements were performed using confocal mode after adding 2 mL of 20 μM CCCP dissovled in complete media.

*Assessment of $T_1$ relaxation time of FNDs in different media:* Ensemble FNDs were deposited onto a petri dish through ten successive applications of 1 μL of a 1 mg mL$^{-1}$ solution, followed by drying under an infrared lamp. $T_1$ relaxation measurements were conducted with wide-field mode after the addition of 2 mL of respective test solutions (water, MES buffer, PBS, or 1 M NaCl). A 2-hour equilibration period was implemented prior to data acquisition to ensure full environmental stabilization of the FNDs.

**Acknowledgements**
We thank Chunying Yin for her helpful with cell sample preparation for electron microscopy and Qinsong Hu for his helpful with cell viability experiment. This work was supported by the National Natural Science Foundation of China (Grant No. T2125011), the Chinese Academy of Sciences (Grant No. YSBR-068), Innovation Program for Quantum Science and Technology (Grant No. 2021ZD0302200, 2021ZD0303204), New Cornerstone Science Foundation through the XPLORER PRIZE, and the Fundamental Research Funds for the Central Universities.

**Conflict of Interest**
The authors declare no conflict of interest.

**Data Availability Statement**
((include as appropriate, including link to repository))